\documentclass[journal]{IEEEtran}

\usepackage{ifpdf}
\usepackage{cite}
\usepackage{amsmath,amssymb,amsfonts}
\usepackage{algorithmic}
\usepackage{graphicx, adjustbox}
\usepackage{array}
\usepackage{enumitem}
\usepackage{breqn}
\usepackage{ dsfont }
\usepackage{ upgreek }
\usepackage{textcomp}
\usepackage{multirow}
\usepackage{algorithm}
\usepackage{algorithmic}
\usepackage{caption}
\usepackage{pgfplots}
\usepackage{tkz-euclide,subfigure}
\usepackage{tikz}
\usepackage{hyperref}
\usepackage{balance}

\DeclareMathOperator*{\minimize}{minimize}

\begin{document}

\title{Secure Computation Offloading in Blockchain based IoT Networks with Deep Reinforcement Learning}

\author{Dinh C. Nguyen,~\IEEEmembership{Member,~IEEE,}
	Pubudu N. Pathirana,~\IEEEmembership{Senior Member,~IEEE,}\\
       Ming Ding,~\IEEEmembership{Senior Member,~IEEE,} and
       Aruna Seneviratne,~\IEEEmembership{Senior Member,~IEEE}
\thanks{*This work was supported by CSIRO Data61, Australia.}
\thanks{Dinh C. Nguyen and Pubudu N. Pathirana are with the School of Engineering, Deakin University, Waurn Ponds, VIC 3216, Australia
(e-mails: cdnguyen@deakin.edu.au, pubudu.pathirana@deakin.edu.au}% <-this % stops a space
\thanks{Ming Ding is with Data61, CSIRO, Australia (email: ming.ding@data61.csiro.au)}% <-this % stops a space
\thanks{Aruna Seneviratne is with the School of Electrical Engineering and Telecommunications, University of New South Wales (UNSW), NSW, Australia (email: a.seneviratne@unsw.edu.au)}
}

\markboth{Accepted at IEEE Transactions on Network Science and Engineering}%
{}

\maketitle

\begin{abstract}
For current and future Internet of Things (IoT) networks, mobile edge-cloud computation offloading (MECCO) has been regarded as a promising means to support delay-sensitive IoT applications. However, offloading mobile tasks to cloud is vulnerable to security issues due to malicious mobile devices (MDs). How to implement offloading to alleviate  computation burdens at MDs while guaranteeing high security in mobile edge cloud is a challenging problem. In this paper, we investigate simultaneously the security and computation offloading problems in a multi-user MECCO system with blockchain. First, to improve the offloading security, we propose a trustworthy access control using blockchain, which can protect cloud resources against illegal offloading behaviours. Then, to tackle the computation management of authorized MDs, we formulate a computation offloading problem by jointly optimizing the offloading decisions, the allocation of computing resource and radio bandwidth, and smart contract usage. This optimization problem aims to minimize the long-term system costs of latency, energy consumption and smart contract fee among all MDs. To solve the proposed offloading problem, we develop an advanced deep reinforcement learning algorithm using a double-dueling Q-network. Evaluation results from real experiments and numerical simulations demonstrate the significant advantages of our scheme over existing approaches. 
\end{abstract}


\IEEEpeerreviewmaketitle

\section{Introduction}
\label{Sec:introduction}
\textcolor{black}{Recent years have witnessed the explosion of mobile technologies with the proliferation of mobile devices (MDs) such as smartphones, tablets, wearable devices, etc, which have been driving the evolution of Internet of Things (IoT). MDs can be used to run IoT applications such as smart home, smart healthcare, and smart city with high flexibility and efficiency. However, due to the rapid growth of mobile data traffic, executing extensive IoT applications merely on overloaded MDs is incapable of providing satisfactory quality of service (QoS) to users  \cite{add2}. Fortunately, with the recent advancement of communication technologies in 5G networks, MDs now can offload the workload of IoT applications (or computation tasks) to a cloud server for execution, which can provide an effective alternative to mitigate the data computation pressure on MDs.  To further improve the efficiency of mobile computation, mobile edge computing (MEC) has emerged as a promising solution to enable MDs to offload their computation tasks to a nearby edge server \cite{2}. In fact, the MEC server is less powerful than a remote cloud, but it is located at the edge of the network,  with a close proximity to MDs, which enables highly efficient IoT data computation with much lower transmission delay, compared with the remote cloud \cite{3}. }

\textcolor{black}{More interestingly, the combination of cloud and edge computing leads to a new paradigm of mobile edge-cloud computation offloading (MECCO) to facilitate offloading computation for IoT networks \cite{4}. The MECCO model can offer application developers highly effective computing services in the mobile edge-cloud by obtaining the advantages of both edge and cloud computing to satisfy diverse users' QoS requirements. Mobile applications without latency requirements (e.g., big data analysis) can be offloaded to the resourceful cloud, while the others (e.g., smart healthcare, smart home) with time-sensitive requirements can be executed at the edge server for fast-response services. Clearly, the MECCO architecture can offer promising solutions with high flexibility to achieve communication and computation objectives for future IoT applications in 5G networks \cite{5}. }

\textcolor{black}{However, the MECCO also poses many challenges on computation offloading and one of the major challenges is security \cite{6}. As mobile task offloading relies on MDs in a dynamic environment where mobile users are untrusted, the MEECO is vulnerable to various types of threats. Unauthorized MDs may gain malicious access to exploit computing cloud services without the consent of a central authority. Further, attackers can threaten computation resources on the cloud to obtain mobile data, leading to privacy concerns of cloud-based IoT applications. Therefore, how to guarantee security for mobile offloading is critical to any MECCO system. }

Recently, blockchain has emerged as a promising approach to tackle security issues in future IoT networks, including mobile offloading systems \cite{7}, \cite{add1}. The concept of blockchain is based on a peer-to-peer network architecture in which transaction information is distributed among multiple nodes and not controlled by any single centralized entity. The blockchain utilises public-key cryptography to establish an append-only, immutable chain of blocks which are publicly accessible to all blockchain entities in a verifiable and trustworthy manner \textcolor{black}{\cite{8}}. Blockchain with its decentralized and trustworthy nature has been integrated with cloud IoT systems for security guarantees such as secure access control and data management among IoT devices. Specially, smart contract \cite{9}, a self-operating computer program running on the blockchain platform has been demonstrated its feasibility in various IoT security problems. For example, smart contracts were employed to design an access control mechanism that was capable of tracking data exchanges between untrusted IoT devices with the ability to detect malicious access while providing provenance and auditing on mobile data \cite{10}. Further, blockchain and smart contracts were adopted to offer access control solutions to manage and protect cloud resources among cloud nodes \cite{11}. With such security capabilities, it is believed that blockchain and smart contracts can be applied in mobile cloud IoT contexts, especially in MECCO systems, to fulfill security objectives for mobile task offloading.

\subsection{Related Works}
Many works were proposed to investigate computation offloading issues with edge-cloud computing in IoT networks with blockchain. The offloading strategies proposed in \cite{12}, \cite{13} utilized edge or cloud services to tackle blockchain-based IoT computation issues with the objective of minimizing offloading costs and cloud resources by leveraging Lyapunov or conventional convex optimization methods. But such conventional offloading optimization algorithms only work well for low-complexity online models and usually require prior knowledge of system statistics that is difficult to acquire in practical scenarios. To overcome such challenges, Reinforcement Learning (RL) \cite{14} has emerged as an efficient technique which allows a learning agent to adjust its policy and derive an optimal solution via trial and error to achieve the best long-term goal without requiring any prior environment information. Nevertheless, in complex offloading problems with multi-user multi-IoT device scenarios, the dimension of state and action space can be extremely high that makes RL-based solutions inefficient. Fortunately, Deep Reinforcement Learning (DRL) methods \cite{add2} such as deep Q-network (DQN) have been introduced as a strong alternative to solve such high-dimensional problems in blockchain-based edge cloud offloading tasks \cite{15}, \cite{16}. The authors in \cite{23} paid attention to an edge offloading scheme for blockchain mining tasks that can be offloaded to the edge clouds, in order to enhance the quality of service (QoS) and mitigate the mining burden posed on mobile miners. The study in \cite{24}  proposed a blockchain-based MEC model for future wireless networks, where computation offloading and resource allocation are jointly optimized with respect to spectrum allocation, block size, number of consecutive blocks via a double-dueling deep Q network. The authors in \cite{25}   paid attention to a cooperative computation offloading framework for blockchain-based IoT networks. An MA-DRL algorithm is designed which allows IoT devices to collaboratively explore the offloading environments in order to minimize long-term offloading costs. 

Moreover, the security of task offloading in edge cloud computing has been also explored recently by using blockchain and smart contracts. For example, the authors in \cite{17} leverage blockchain to create a secure communication protocol with group signatures and covert channel authorization techniques to guarantee the validity of users in edge computing. A blockchain-empowered framework is also proposed in \cite{18} for implementing resource trading and task assignment in edge computing as the smart contracts which can provide reliable resource transactions and immutable records in the blockchain. Smart contracts have been also adopted in \cite{19} for facilitating a vehicular edge consortium blockchain, which enables secure resource sharing while motivating vehicles to share their computation resources with service requesters. To support security and privacy for cognitive edge computing, spatio-temporal smart contracts are provided in \cite{20} with incentive mechanisms to accelerate the economy sharing in smart cities.

\subsection{Main Contributions and Paper Structure}
\textcolor{black}{Despite these research efforts, there has been little attention given to the design of security, e.g., access control for task offloading in these existing works \cite{15,16, 23,24,25}.  Moreover, the integration of edge and cloud computing in blockchain has not been investigated fully for IoT computation offloading scenarios \cite{18,19,20}. Motivated by such limitations, in this paper, we propose a novel secure computation offloading model for IoT networks on mobile edge-cloud based on blockchain and DRL techniques. In particular, we provide a comprehensive solution to the access control and efficient computation problems of the MECCO system on blockchain to satisfy both the security and offloading requirements. Also, different from existing works, we evaluate the proposed scheme by conducting both real experiments and numerical simulations. In a nutshell, the main contributions of this paper are highlighted as follows:
\begin{enumerate}
	\item We consider a new secure computation offloading scheme for mobile edge cloud computing with blockchain where MDs can offload their IoT data tasks to the cloud or edge server for computation under an access control mechanism for security guarantees. 
	\item	We propose a trustworthy access control mechanism specific to edge task offloading by using a trustworthy smart contract on blockchain. The main purpose of our access control design is to perform user authentication, offloading verification, and manage offloaded mobile data, which thus provides high security for the MECCO system. 
	\item	We formulate an optimization problem by taking both computation offloading and smart contracts into account, which has not been considered in the open literature. This is enabled by the joint consideration of the offloading decisions, the allocation of computing resource and radio bandwidth, and smart contract usage.	To this end, we proposed an advanced DRL algorithm to obtain the optimal offloading policies for all MDs. 
	\item	We investigate the effectiveness of the proposed MECCO framework in terms of access control and offloading performances by conducting both real experiments and numerical simulations. The implementation results demonstrate that the proposed scheme not only provides reliable authentication for task offloading but also reduces offloading costs, compared to the existing schemes. 
\end{enumerate}}

The remainder of this paper is organized as follows. Section~\ref{Sec:Architecture} presents the integrated edge and cloud architecture with blockchain for secure computation offloading. In Section~\ref{Sec:SystemModel}, the system models for both access control and computation offloading are described. Then, we formulate the computation offloading, edge resource allocation, radio bandwidth resource, and smart contract cost as a joint optimization problem. In Section~\ref{Sec:Solution}, we propose and access control mechanism with smart contracts and computation offloading approaches with an advanced DRL algorithm for our MECCO system. The performances of the proposed MECCO system are investigated in Section~\ref{Sec:Evaluation} by conducting both real experiments to evaluate the access control performance and numerical simulations. Finally, conclusions are given in Section~\ref{Sec:Conclude}.

\section{Network Architecture}
\label{Sec:Architecture}
%In this section, we first present the integrated edge and cloud architecture on a blockchain platform. We describe the key concept of the proposed MECCO system which consists of a joint access control and computation offloading scheme. Then, we explain the key benefits brought by blockchain for secure task offloading in edge cloud computing. 
\subsection{Mobile Edge Cloud for Blockchain-IoT Networks}
\begin{figure}
	\centering
	\includegraphics[width=0.95\linewidth]{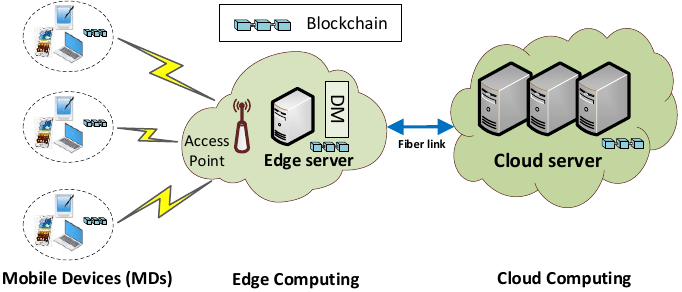}
	\caption{The proposed mobile edge-cloud network architecture with blockchain for secure mobile offloading.  }
	\label{Fig:Overview}
\end{figure}

We propose an integrated edge-cloud architecture for IoT networks with blockchain as shown in Fig.~\ref{Fig:Overview}, including three main layers, namely mobile devices layer, edge computing layer and cloud computing layer. The key features of each layer are presented as follows.

\subsubsection{Mobile Devices Layer} This layer consists of a network of MDs, such as smartphones, tablets, and sensor devices. Such MDs are connected together in the IoT network by the blockchain. Each MD has a blockchain account to join into the network for various functionalities such as collecting data or performing task offloading to cloud servers. For example, in our MECCO scenario, each MD needs to send requests to the cloud server for task offloading. Once the cloud server verifies the request, a response will be returned to the MD for offloading permission so that the device can start to offload its tasks to edge or cloud server for computation. 

\subsubsection{Edge Computing Layer} This layer includes a wireless access point (AP) or base station (BS) for wireless communication with local devices, a Decision Maker (DM) for task execution decision, and a light-weight edge server for instant data processing. This layer can provide low-latency computation services at the edge of the network. However, for complex computation tasks, the edge server needs to forward them to the resourceful cloud server by a wired line to avoid the task overload on the edge layer. In addition, the edge server also acts as a blockchain entity to establish trustworthy communication with cloud nodes and MDs on the blockchain network for security guarantees. Any transactions and offloading activities in the offloading system will be recorded by blockchain and also broadcast to the edge server to achieve a common agreement on offloading management. In this paper, for simplicity, we only consider one edge server in our MECCCO system. However, our model can easily extend to multi-edge scenarios where each edge server server acts as a decentralized blockchain node for system throughput and security improvements.  

\subsubsection{Cloud Computing Layer} This layer includes multiple virtual machines with powerful computation and storage capabilities to solve complex computation tasks from local IoT devices. In our MECCO architecture, the cloud layer also contains a network manager as a blockchain entity to control all user access, an admin for smart contract management and a group of miners for transaction mining. All cloud nodes operate on the blockchain platform in a decentralized and secure manner and link securely to edge server and MDs via the blockchain network. 

\textcolor{black}{\subsection{Motivations of Using Blockchain for Secure Computation Offloading in Edge Cloud Computing}}

\textcolor{black}{Due to the decentralized, immutable, and traceable features, blockchain is able to provide higher security degrees to computation offloading in edge cloud computing, compared to the traditional security techniques \cite{liextended2019}. Indeed, security solutions such as access control based on blockchain not only manage effectively offloading, but also enable reliable authentication on all offloading behaviours to preserve the cloud resources in a decentralized manner. The motivations of using blockchain for security, e.g., access control, in computation offloading are enabled by its unique advantages over the conventional access control solutions \cite{riadsensitive2019, fanefficient2019} which are explained as follows: 
\begin{itemize}
	\item Blockchain can provide a decentralized management solution for the offloading system. IoT data offloaded from MDs can be stored in the peer-to-peer storage in the blockchain network without relying on any central authority, which ensures fast data access and enhances significantly data security of mobile users \cite{17}. 
	\item By incorporating blockchain into the edge-cloud computing network, the offloading system can achieve a trustworthy access control by using smart contracts which enables to authorize automatically devices and distinguish users from adversaries, aiming to prevent malicious offloading behaviours and potential threats to cloud computation resources. Consequently, the  data integrity and offloading validity of the system can be significantly improved \cite{yangauthprivacychain2020}. 
	\item With its decentralized and immutable nature, blockchain can work well in untrusted environments like our considered IoT scenario where there is no need the trust between cloud server, edge server and IoT devices. Particularly, the peer-to-peer network architecture provided by blockchain can achieve robust access control with high data integrity and system security for the mobile offloading \cite{18}. 
\end{itemize}}

\subsection{Description of the Proposed MECCO System }

\begin{figure}
	\centering
	\includegraphics[width=0.98\linewidth]{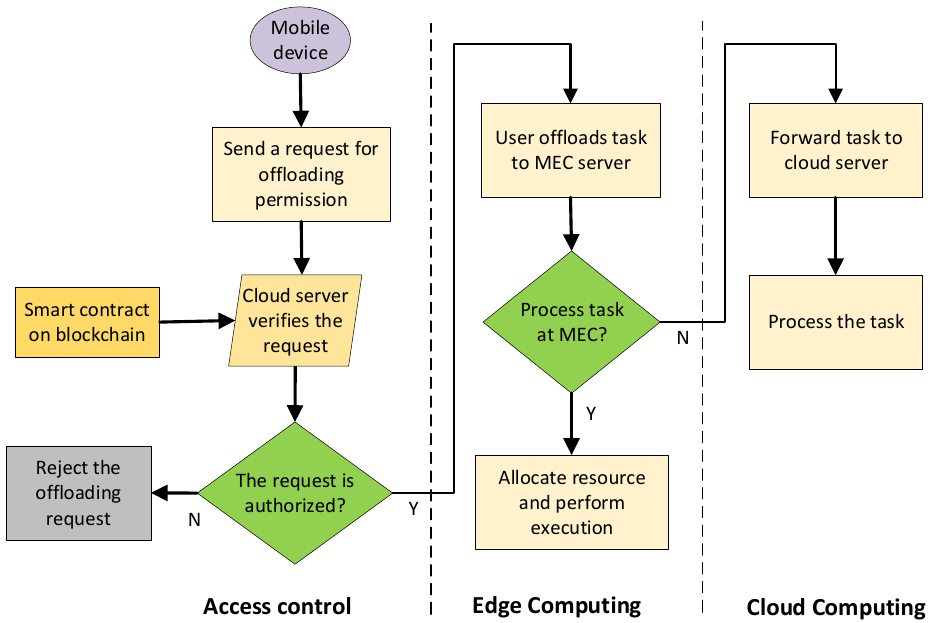}
	\caption{Flowchart of computation task offloading with access control on edge-cloud. }
	\label{Fig:Flowchart}
	\vspace{-0.1in}
\end{figure}

With the network settings, we describe the proposed MECCO system with a focus on access control and offloading concepts on the blockchain network. To guarantee the QoS requirements of mobile users, the combination of edge-cloud computing and blockchain should be considered carefully. In fact, since the dynamic and scalable characteristics of the mobile multi-IoT environment, access control and computation offloading requires a comprehensive design to achieve both security and offloading goals. To provide a complete computation solution for the proposed MECCO system, we introduce a network workflow as indicated in Fig.~\ref{Fig:Flowchart}.  The procedure includes two phases: access control and computation offloading, which are described as follows. 

First, a MD initializes a request as a blockchain transaction to start the computation offloading process and sends this request to the cloud server via the wireless access point (Note that for convenient management, access control for all MDs is implemented at the central cloud server). Then, the cloud server will authorize this request using an access control mechanism enabled by smart contracts. Based on predefined strict control policies, smart contract will identify, analyse and make decisions to accept or refuse the request. If the request is authorized successfully, a response will be returned to the MD so that the device can offload its tasks. Now the access control process finishes and the transaction will be recorded and stored on the blockchain network in a secure manner. In the second phase, the authorized MD will choose to offload its computation tasks to the edge or cloud server for calculation. At each offloading period, based on QoS requirements and current network conditions (task size, available edge resource, channel bandwidth resource), the decision maker (DM) at the edge layer \cite{15} will perform optimization to decide where the mobile task should be executed, i.e., in the MEC server or cloud server, for optimal computation benefits (e.g., minimum offloading costs). If the tasks are executed at the edge layer, the MEC server needs to allocate communication resource as well as computation resource to each MD. However, if the tasks exceed the computing capability of the MEC server, such tasks should be forwarded to the resourceful cloud server for calculation. Note that data offloaded from MDs can be stored securely in a decentralized cloud storage on blockchain [15]. The data storage management is beyond the scope of this paper, and details can be referred to our previous work [15].

%Clearly, in our MECCO settings, the combination of edge and cloud computing on blockchain enables MDs to perform cooperatively their tasks with optimal computation costs, while security of offloading process is guaranteed. Design of secure task offloading schemes will be presented in the next sections. 
\section{System Model and Offloading Problem Formulation}
\label{Sec:SystemModel}
%In this paper, we consider a MECCO system on blockchain as illustrated in Fig.~\ref{Fig:Overview}, which consists of a remote cloud server, a MEC server and a network of MDs. All MDs are connected to the edge and cloud server via a wireless access point (AP). The proposed MECCO system includes two schemes, access control and computation offloading scheme. A MD first should be verified by an access control mechanism to obtain an offloading right. Then, the authorized MD can offload its task to the server for computation. Details of these two models are presented in the following subsections. 
The proposed MECCO system includes two schemes, access control and computation offloading, which will be presented in the following. 
\subsection{Access Control Model}

We propose an access control model on a blockchain network for the MECCO system as shown in Fig.~\ref{Fig:AccessControl}. For better offloading management, access control for all MDs is implemented at the cloud server as explained in the previous section, and therefore the MEC server is ignored in the access control scheme. 

%Note that in our design, blockchain utilizes community validation to achieve synchronization on distributed ledgers which are replicated across cloud nodes (MECCO manager, admin and miners) within the peer cloud network. This decentralized cloud architecture with distributed access control gets rid of the single-point failure problems of traditional centralized cloud system, which ensures high system availability and computation service performances. In the following, we introduce the key components and then explain the operation concept of the proposed access control scheme.
\subsubsection{Key Components of The Access Control Scheme}
\begin{figure}
	\centering
	\includegraphics[width=0.98\linewidth]{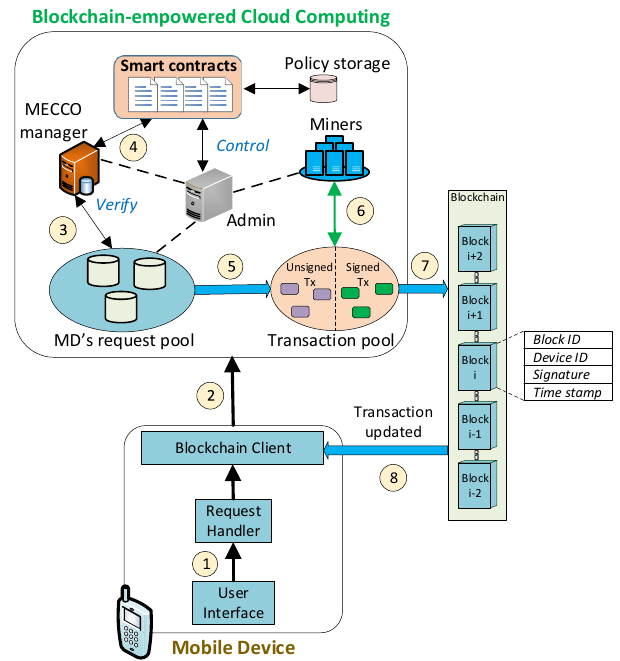}
	\caption{Workflow of blockchain-based access control scheme for the MECCO system. }
	\label{Fig:AccessControl}
	\vspace{-0.17in}
\end{figure}
As shown in Fig.~\ref{Fig:AccessControl}, the access control scheme consists of four main components: MECCO manager, admin, smart contracts and miners.
\begin{itemize}
	\item \textit{MECCO Manager:} The MECCO manager plays a significant role in our access control framework. It is responsible to monitor all offloading events on the blockchain network, including offloading requests and access authentication for MDs. The management capability of MECCO manager is enabled by smart contracts through strict user policies. 
	\item \textit{Admin: }It is used to manage transactions and operations on cloud by the means of adding, changing or revoking access permissions. Admin is responsible to deploy smart contracts and the only entity with the ability to update or modify policies in smart contracts.
	\item \textit{Smart Contracts: }The smart contracts define all operations allowed in the access control system. MDs can interact with smart contracts by the contract address and Application Binary Interface (ABI). Smart contracts can identify, validate request and grant access permissions for MDs by triggering transactions or messages. The smart contract and its operations are accessible to all blockchain entities. It is considered as core software in our access control scheme.
	\item \textit{Miners: }\textcolor{black}{We employ a group of virtual machines on the cloud as miners that perform mining tasks to validate the data blocks consisting of transactions of MDs via a consensus mechanism such as Proof-of-Work (PoW) \cite{30}. Here, the fastest miner which solves the computational puzzle is rewarded for its mining contribution, and it also verifies the data block that is then sent along with the signature to other miners for validation. If all miners achieve a consensus, the validated block is then appended into the blockchain in a chronological order. }
	
\end{itemize}
\subsubsection{Access Control Concept for Secure Computation Offloading}
%In this subsection, we describe the operation concept of access control for secure computation offloading in the MECCO system based on smart contracts and blockchain. The use of smart contract allows managing automatically transactions and operations on blockchain with high efficiency and reliability. Specially, at each MD, we design a blockchain client module, which implements a full functionality to participate in the cloud blockchain network. This module is responsible to encode transactions and data requests, sign digitally on transactions and connect with blockchain for transaction tracking. Besides, the user interface module is designed for user interaction, and a request handler module for processing user request information. 

The workflow of access control on blockchain is illustrated in Fig.~\ref{Fig:AccessControl}. A description of each step is provided as follows.

\begin{enumerate}[label=\large\protect\textcircled{\small\arabic*}]
\item A MD initializes a request as an offloading transaction for offloading computation tasks to the edge-cloud server.
\item The blockchain client processes and sends the request to the storage pool so that the MECCO manager and smart contracts can verify. 
\item  The MECCO manager collects the requests of MDs in the storage pool based in a first-come-first-served manner. 
\item 	The MECCO manager verifies the request by smart contracts with a strict control policy. If the request is accepted, a response will be returned to the MD for offloading data.
\item 	Offloading transactions are grouped into data blocks, which are then inserted into the transaction pool for confirmation by miners.
\item 	The miners validate the data blocks and sign them with digital signature to append to the blockchain.
\item 	The offloading transaction is added to the blockchain network and broadcast to all MDs within the MECCO system. 
\item 	The offloading transaction is updated at the MD for tracking via the blockchain client. 
\end{enumerate}
The access control process will be explained in the following steps:

\begin{itemize}
	\item \textit{Step 1-Initialization (executed by the mobile gateway):} The MD needs to create a blockchain account to join the blockchain network and initializes a request as a transaction for offloading computation tasks. The MD uses the blockchain client module to interact with blockchain on clouds. In the task offloading, this module will create a storage transaction $T_s$ as a request with information index including request metadata (Device ID), digital signature and timestamp for verification. The request then will be sent to the MECCO manager on cloud for verification via a wireless network. (Steps 1, 2 in Fig.~\ref{Fig:AccessControl}).
	\item \textit{Step 2-Verification of the offloading request (executed by the MECCO manager):} After receiving a transaction from the MD, the MECCO manager will issue a signal to smart contract as a notification of a new task offloading request. By using the policy list in the policy storage, the smart contract can verify the transaction and if it is accepted, a message will be returned to the MD via the blockchain client to allow to offload computing tasks (Step 3, 4 in Fig.~\ref{Fig:AccessControl}). 
	\textcolor{black}{Note that in the practical limited storage, a threshold can be set up at the MECCO manager to manage the request flow from the MDs. In this regard, only a certain number of requests under the pre-defined threshold are processed by the MECCO manager for authentication in a time period, while the future requests will queue to be handled in the next time window, where queue theories may be useful to model this process.}
	\item \textit{Step 3-Adding the offloading transaction to blockchain (executed by miners):} After verification, metadata of offloading transactions (Device ID as shown in Fig.~\ref{Fig:AccessControl}) is also inserted into the unsigned transaction pool. The miners will form periodically transactions in the pool into blocks for mining (Step 5 in Fig.~\ref{Fig:AccessControl}). The fastest miner which verifies the data block will send the signature to other miners for validation. If all miners achieve an agreement, the validated block with its signature is then appended to the blockchain in a chronological order. Finally, all network MDs receive this block and synchronize the copy of the blockchain via the blockchain client. The workflow for the above process is shown in Steps 6, 7, 8 in Fig.~\ref{Fig:AccessControl}. 
\end{itemize}

%In summary, the proposed access control scheme can authorize effectively offloading behaviours of MDs by using smart contracts. Our approach therefore can prevent malicious access to the cloud resources, which improves significantly security for our MECCO network. MDs which are authorized by the access control mechanism now can offload their computation tasks to edge or cloud server for calculation. In the next subsection, we will formulate the task offloading problem for our proposed MECCO system. 

\subsection{Computation Offloading Model}

In the computation offloading model, we assume that MDs are authorized by our access control scheme as designed in the previous subsection. We consider realistic IoT applications (e.g., speech recognition or big data analysis) where computation tasks can be very large and thus inefficient to be processed locally by MDs. Therefore, the authorized MDs will have to offload their tasks to edge or cloud server for efficient execution. In this subsection, we propose a new offloading scheme for our MECCO scenario. We first formulate the task model and computation model, and then we describe the problem formulation in details.

\subsubsection{Task Model}
We consider a task model as shown in Fig.~\ref{Fig:Overview}. We denote a set of MDs as $\mathcal{N} = \{1,2,..., N\}$. It is assumed that each MD has a computation task to be completed. For each MD $n$, the computation task can be formulated as a variable tuple $R_n=(D_n, X_n, \tau_n )$. Here $D_n$ (in bits) denotes the data size of computation task of the MD $n$. We also assume that the size of $D_n$ is fixed when it is offloaded to edge or cloud server. $X_n$ (in CPU cycles/bit) denotes the total number of CPU cycles required to accomplish the computation for the task $R_n$. Moreover, $\tau_n$ (in seconds) reflects the maximum tolerable delay of task $R_n$. Note that the information of $D_n$ and $X_n$ can be obtained by using program profilers \cite{16}. In this paper, we focus on a joint optimization problem of offloading decision, edge computation resource and bandwidth resource, which has been not studied well in the literature studies. 

We define an offloading decision vector as $A =[\alpha_1, \alpha_2,..., \alpha_N]$ where $\alpha_n = \{\alpha_n^e, \alpha_n^c \}$. Here $\alpha_n^e, \alpha_n^c \in \{0,1\}$  and $\alpha_n^e + \alpha_n^c =1$. If the task is offloaded to the edge or cloud server, the corresponding parameter is 1, otherwise it is 0. Note that each task is executed at only a platform at each offloading period. %For tractable network analysis, similar to the existing works \cite{15,16,17}, we assume that all MDs and wireless network remain stationary during the offloading period. This assumption is feasible to many IoT applications such as speech recognition or face recognition in our concerned scenarios, where the offloading period is much shorter than the timescales of MD mobility and wireless network dynamics. 
Further, we also consider the resource allocation problem for computation offloading. For edge computing, the MEC server needs to allocate its computing resources to each MD to perform task execution. For cloud computing, the task needs to be offloaded by the MD to the edge layer and then is forwarded to the remote cloud via a wired link. Due to the powerful computational capacity of the cloud server, the problem of cloud resource allocation is ignored in our paper. Nevertheless, the allocation of limited radio bandwidth should be considered to improve offloading efficiency of our MECCO system. Assuming that the total radio bandwidth of our MECCO system is $B$ Hz, we allocate part of the bandwidth resource to each MD to avoid interference between them \cite{25}. We normalize the assigned bandwidth to the MD $n$ as $w_n \in [0,1]$, then we have $\sum_{1}^{N} w_n = 1$. We also denote the channel gain between MDs and the MEC server as $h_n$, then the transmission data rate of the MD $n$ can be calculated as
\begin{equation}
r_n = w_nBlog_2(1+\frac{p_nh_n}{w_nN_0B}),
\end{equation}
where $p_n$ is the transmit power (W) of the MD $n$, $N_0$ is additive noisy power spectral density (dBm/Hz). 

In the following, we formulate the computation offloading model of edge computing and cloud computing with a focus on computation latency and energy consumption analysis for our MECCO system. 

\subsubsection{Computation Model}
The data tasks of MDs can been offloaded to the edge or cloud server for execution. 

\textit{2.1) Edge Computing: }We consider the case when the task $R_n$ of the MD $n$ is offloaded to the MEC server for computation ($\alpha_n^e=1$). We denote $T_n^e$ as the edge computing latency which includes the transmission delay for the MD $n$ sending data to the MEC server and the execution time on the MEC server. Similar to \cite{24}, the total edge computing latency can be expressed as 
\begin{equation}
T_n^e =  \frac{D_n}{r_n} + \frac{X_n}{f_n^{e}},
\end{equation}
\\
where $f_n^{e}$ (in CPU cycles/s) denotes the edge computation resource allocated to the MD $n$. Note that edge resource allocated to all MDs should not exceed the total computational capacity of the MEC server $\sum_{n=1}^{N} f_n^e \leq F^e$. 

Moreover, the energy cost for offloading to the MEC server consists of energy consumption for transmitting data and execution. Denote $p_n^i$ as the power consumption (in watt) of the MD $n$ in idle status, the total energy consumption of offloading data to the MEC server is given as
\begin{equation}
E_n^e =  \frac{p_nD_n}{r_n} + \frac{p_n^iX_n}{f_n^{e}}.
\end{equation}

\textit{2.2) Cloud Computing: }If the computation task is offloaded to the cloud server ($\alpha_n^c=1$), the MD $n$ needs to offload its task to the MEC server which then forwards it to the remote cloud server for computation via a wired link. The cloud server also allocates its computation resources to compute the task efficiently. We denote the data rate of the wired link for transmitting the task of the MD $n$ as $r_n^w$ and the cloud resource allocated to the MD $n$ as $f_n^c$. Then then the total cloud computing latency can be expressed as \cite{24}
\begin{equation}
T_n^c =  \frac{D_n}{r_n} + \frac{D_n}{r_n^w} + \frac{X_n}{f_n^{c}}.
\end{equation}

Besides, offloading data to the cloud server also incurs the energy cost, which can be calculated as 

\begin{equation}
E_n^c =  \frac{p_nD_n}{r_n} + p_n^i(\frac{D_n}{r_n^w} + \frac{X_n}{f_n^{c}}).
\end{equation}

According to (2)-(5), the computation latency and energy consumption of the MD $n$ in our MECCO system can be expressed respectively as

\begin{equation}
T_n= T_n^e +T_n^c,
\end{equation}
\begin{equation}
E_n = E_n^e +E_n^c.
\end{equation}

\subsubsection{Smart Contract Cost for Offloading Authentication}
 We also consider the contract execution cost for offloading authentication which has not been explored in previous blockchain-based offloading works \cite{13,17,18}. In the cloud blockchain, e.g., Ethereum, all transactions have fees which are measured in units of gas, which can be regarded as a metric to standardize the contract cost \cite{21}. To execute a smart contract, a MD's account has to pay a certain amount of gas using Ether, a common cryptocurrency in Ethereum, to specify the amount of computation and storage required by a transaction. Each offloading transaction in cloud Ethereum must specify a gas limit value $\varpi_n$, which is the maximum amount of gas for executing a transaction. A transaction whose gas cost is beyond the current block gas limit will be rejected by the network. Transactions also determine a gas price $\xi$ which is the rate paid to cloud miners in Ether per unit of gas. We denote $g_n$ to present the amount of gas used when executing the access control contract. Accordingly, the contract fee (in Ether) that the authorized MD $n$ needs to pay for authentication as \cite{22}: 
\begin{equation}
C^{sc}_n =  \xi*min\{g_n, \varpi_n\}. 
\end{equation}
The key notations used in this paper are listed in Table~\ref{table}. 
\begin{table}
	\centering
	\caption{List of key notations. }
	\label{table}
	
	\setlength{\tabcolsep}{5pt}
	\begin{tabular}{|c|p{6cm}|}
		\hline
		\textbf{Notation}& 
		\textbf{Description}
		\\
		\hline
		$\mathcal{N}, \mathcal{N}^'$& Set of all MDs/ authorized MDs in MECCO system
		\\
		\hline
		$D_n$&	The data size of computation task of MD $n$ 
		\\
		\hline
		$X_n$&	The total number of CPU cycles per task
		\\
		\hline
		$\tau_n$&	Completion deadline for executing a task
		\\
		\hline
		$\alpha_n^e$, $\alpha_n^c$& Process the task at edge/cloud server
		\\
		\hline
		$p_n, p_n^i$& The transmit power/idle power of the MD $n$ 
		\\
		\hline
		$N_0$& Additive noisy power spectral density
		\\
		\hline
		$B$& The total radio bandwidth of our MECCO system
		\\
		\hline
		$w_n$& The allocated bandwidth to the MD $n$
		\\
		\hline
		$h_n$& The channel gain between MDs and MEC server
		\\
		\hline
		$r_n$& The transmission data rate of MD $n$
		\\
		\hline
		$f_n^e$, $f_n^c$& The computational capacity of edge/cloud server
		\\
		\hline
		$F^e$& The total MEC computational capacity
		\\
		\hline
		$T_n^e$, $T_n^c$&  Computation latency of edge/cloud processing
		\\
		\hline
		$E_n^e$, $E_n^c$&  Energy consumption of edge/cloud processing
		\\
		\hline
		$C_n^{offload}$& The offloading cost of MECCO system
		\\
		\hline
		$g_n$ 	& The gas fee of smart contract execution
		\\
		\hline
		$C^{sc}_n$& The offloading authentication cost 
		\\
		\hline  
	\end{tabular}
	\label{tab1}
	\vspace{-0.16in}
\end{table}

\subsection{Formulation of the Secure Offloading Problem }

In this subsection, we formulate the computation offloading, edge resource allocation, radio bandwidth resource, and smart contract cost as a joint optimization problem. Our objective is to minimize the sum cost of computation latency, energy consumption, and smart contract cost for all MDs in our blockchain-based MECCO system. We formulate the offloading cost function of the MD $n$ as the weighted sum of computation latency and energy consumption, which is given as 
\begin{equation}
C_n^{offload} = \beta^tT_n +\beta^eE_n,
\end{equation}
where $\beta_n^t, \beta_n^e \in [0,1]$ ($n \in \mathcal{N}$) denote the weight of latency and energy consumption, respectively. Thus, the total cost can be defined as the sum of the smart contract cost $C^{sc}_n$ and the offloading cost $C_n^{offload}$. To this end, we formulate the joint optimization of secure task offloading for the multi-user MECCO, subject to the offloading decisions $\textbf{A} =[\alpha_1^e, \alpha_1^c,..., \alpha_N^e, \alpha_N^c]$, edge resource allocation $\textbf{f} = [f_1, f_2,...,f_N]$, radio bandwidth allocation $\textbf{w} = [w_1, w_2,...,w_N]$, and consumed gas fee  $\textbf{g} = [g_1, g_2,...,g_N]$ as follows: 

\begin{equation*}
\begin{aligned}
& (P1):   \underset{\textbf{A},\textbf{f},\textbf{w}, \textbf{g}}{{\minimize} }
&& \sum_{n=1}^{N} C_n^{offload} + C^{sc}_n\\
& \text{subject to}
&&(C1): \alpha_n^e, \alpha_n^c \in \{0,1\}, \forall \textit{n} \in \mathcal{N},\\
&&&(C2): \alpha_n^e + \alpha_n^c =1, \forall \textit{n} \in \mathcal{N},\\
&&&(C3): \sum_{n=1}^{N} f_n^e \leq F^e, \\
&&&(C4): f_n^e \geq 0, \forall \textit{n} \in \mathcal{N},\\
&&&(C5): 0 < w_n \leq 1, \forall \textit{n} \in \mathcal{N},  \\
&&&(C6): \sum_{n=1}^{N} w_n \leq 1, \\
&&&(C7): T_n \leq \tau_n, \forall \textit{n} \in \mathcal{N}, \\
&&&(C8): g_n \leq \varpi_n.
\end{aligned}
\end{equation*}

Here, the constraint (C1) and (C2) represent the binary offloading decision policy of the MD $n$, offloading to the MEC server or offloading to the cloud server. (C3) and (C4) indicate that the allocated edge resources should not exceed the total computing capacity of the MEC server, while (C5) and (C6) are the constraints of bandwidth allocation. Further, the execution time to complete a computation task should not exceed a maximum time latency value, which is expressed in the constraint (C7). Also, (C8) represents the constraint of gas execution cost in the offloading authentication. It is worth noting that the optimization problem $(P1)$ is not convex due to the non-convexity of its feasible set and objective function with the binary variable \textbf{A}. The size of the problem $(P1)$ can be very large when the number of MDs in MECCO system increases rapidly. \textcolor{black}{ Moreover, traditional optimizations methods such as ADMM cannot solve well the proposed problem with high network dynamics. Indeed, the authors assume that network statistics such as computation resource and radio bandwidth are known before making offloading decisions, which may not be met in highly dynamic blockchain-based MEC environments like our considered scenario. Moreover, the computation offloading policy designs in these works are mostly based on one-shot optimization and fail to characterize the long-term computation offloading performance in real-time task execution. Therefore, we here propose to use a DRL algorithm to adjust dynamically how much edge computation and bandwidth resources should be allocated to a certain MD based on MDs task size so as to achieve the optimal offloading for the MECCO system, which at the same time does not need a priori knowledge of network statistics.}

\section{	Proposed Solutions}
\label{Sec:Solution}
In this section, we propose access control and computation offloading approaches for our MECCO system. 
\subsection{	Access Control for MECCO System with Smart Contracts}
In this subsection, we design a smart contract to formulate
our access control scheme. We also provide an access protocol
that presents the workflow of access control for the MECCO system. 
\subsubsection{Smart Contract Design}
We first create an \textit{AccessControl} contract controlled by the admin to monitor transaction operations in our MECCO network on blockchain. Here we use an Ethereum blockchain platform \cite{30} due to its adaptable and flexible features, which allow to build any blockchain applications such as our IoT scenario. We denote $PK$ as the public key of MD. The contract mainly provides the following four functions.
\begin{itemize}
	\item \textit{AddMD(PK): (executed by Admin)} This function allows to add a new MD to the smart contract. MD is identified by their public key and is added into the contract with a corresponding role based on their request. MD information is also kept in cloud storage as part of system database. 
	\item \textit{DeleteMD(PK): (executed by Admin)} It is used to remove MDs from the network based on the corresponding public key. All device information is also deleted from cloud storage. 
	\item \textit{PolicyList(PK): (executed by Admin)} The policy list contains the public keys of all MDs for identification when the smart contract processes new transactions.
	\item \textit{Penalty(PK, action): (executed by Admin)} When detecting an unauthorized offloading request to cloud, the MECCO manager will inform smart contract to issue a penalty to the requester. In our paper, we give a warning message as a penalty to the unauthorized MDs. 
	
\end{itemize}

\begin{figure}
	\centering
	\includegraphics[width=0.98\linewidth]{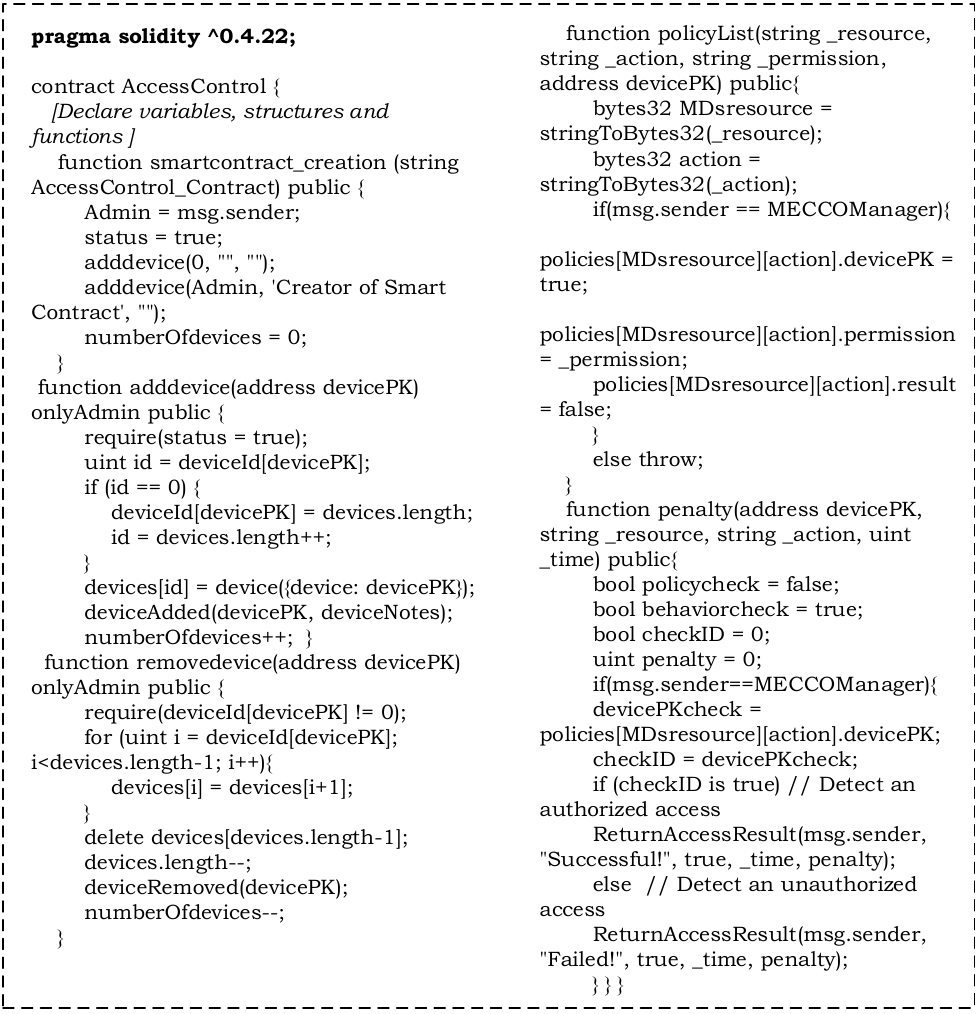}
	\caption{ Pseudo-code of smart contract implementation for access control.  }
	\label{Fig:CodeSC}
	\vspace{-0.17in}
\end{figure}

The smart contract design for the proposed access control scheme on Ethereum blockchain can be seen in Fig.~\ref{Fig:CodeSC}.

\subsubsection{Access Control Protocol}

To operate the access control for computation offloading, we also develop an access control protocol which is performed when a MD executes a transaction for a request of offloading data to the edge-cloud. The access control protocol includes two phases: \textit{Transaction pre-processing} (executed by the MECCO manager) and \textit{Verification} (executed by the Admin). In the first phase, the MECCO manager receives a new transaction \textit{Tx} from a MD. The MECCO manager will obtain the public key $PK$ of the requester by using the \textit{Tx.getSenderPublicKey()} function and send it to the contract for validation. In the next phase, after receiving a transaction with a MD $PK$ from MECCO manager \textit{(msg.sender = ME)}, the admin will verify access rights of the requester based on its \textit{PK} in the policy list of the smart contract. If the \textit{PK}  is available in the list, the request is accepted and now a task offloading permission is granted to the requester. Otherwise, the smart contract will issue a penalty to this request through the \textit{Penalty()} function. In this case, all offloading activities are denied and the request is discarded from the blockchain network. The access control protocol is summarized in Algorithm~\ref{Alg:AccessCOntrol}.

\begin{algorithm}
	\footnotesize
	\caption{Access control for computation offloading}
	\begin{algorithmic}[1]
		\STATE \textbf{Input:} $Tx$ (The offloading request on blockchain)
		\STATE \textbf{Output:} $Result$  (Access result for offloading request)
		\STATE \textbf{Initialization:} \textit{(by the MECCO Manager)} 
		\STATE Receive a new transaction $Tx$ from a MD
		\STATE Get the public key of the requester: $PK\leftarrow Tx.getSenderPublicKey() $
		\STATE Send the public key to Admin ($msg.sender = MECCO manager$)
		\STATE \textbf{Pre-processing the request} \textit{(by Admin)} 
		\IF {$PK$ is available in the policy list}
			\STATE  $policyList(PK) \leftarrow true$
		\ENDIF
		\STATE Decode the transaction $decodedTx \leftarrow abiDecoder.decodeMethod(Tx) $
		\STATE Specify request information: $Addr \leftarrow web3.eth.getData(decodedTx([DataIndex])$
		\STATE Specify \textit{DeviceID}: $D_{ID} \leftarrow Addr(Index[D_{ID}])$; 
		\STATE \textbf{Verification} \textit{(by the smart contract)} 
		\WHILE {true}
			\IF {$policyList(PK) \rightarrow true$}
				\IF {$policyList(D_{ID}) \rightarrow true$}
					\STATE $Result \leftarrow Penalty(PK,"Successful!")$
					\STATE break;
				\ELSE
					\STATE $Result \leftarrow Penalty(PK,"Failed")$
					\STATE break;
				\ENDIF
			\ELSE
				\STATE $Result \leftarrow Penalty(PK,"Failed")$
				\STATE break;
			\ENDIF
		\ENDWHILE
	\end{algorithmic}
\label{Alg:AccessCOntrol}
\end{algorithm}

\subsection{Secure Computation Offloading for MECCO System with Advanced DRL}

\subsubsection{Problem Formulation}
We focus on formulating the secure computation offloading problem via DRL, aiming to minimize the sum cost of all MDs in terms of computation latency, energy consumption, and smart contract cost in the proposed MECCO. We first introduce the offloading framework using DRL where state space, action space and reward are defined.

\begin{itemize}
	\item \textbf{State: } The system state is chosen as $s= \{tc, ec, bw \}$ where \textit{tc} is the total offloading cost of MECCO system ($tc = C$), \textit{ec} is the available computation resource of the MEC server ($ec = F^e - \sum_{n=1}^{N} f_n^e$). Further, \textit{bw} denotes the available bandwidth resource of the MECCO system, and can be specified as $bw= B-\sum_{n=1}^{N} w_n$. Moreover, we estimate the consumed gas cost $g^n$ as the state of smart contract usage when offloading the task of the MD $n$. 
	
	\item \textbf{Action: }The action space is formulated as the offloading decision vector $\textbf{A} =[\alpha_1^e, \alpha_1^c,..., \alpha_N^e, \alpha_N^c]$, edge resource allocation $\textbf{f} = [f_1, f_2,...,f_N]$ and radio bandwidth allocation $\textbf{w} = [w_1, w_2,...,w_N]$. Thus, the action vector can be expressed as $a = [\alpha_1^e, \alpha_1^c, f_1,w_1,...,\alpha_N^e, \alpha_N^c, f_N,w_N]$.
	
	\item \textbf{Reward: }The objective of the RL agent is to find an optimal offloading action $a$ at each state $s$ with the aim of minimizing the sum cost $C(s,a)$ of the offloading cost $C^{offload}(s,a)$ and smart contract cost $C^{sc}(s,a)$ in the MECCO system. Also, the reward function should be negatively related to the objective function of the optimization problem $(P1)$ in the previous section. Accordingly, we can formulate the system reward as
	\begin{equation}
	r(s,a) = -C(s,a) = -(C^{offload}(s,a) + C^{sc}(s,a)).
	\end{equation} 

\end{itemize}
\subsubsection{Basics of Deep Reinforcement Learning}
The principle of Reinforcement Learning (RL) can be described as a Markov Decision Process (MDP) \cite{16}. In the RL model, an agent can make optimal actions by interacting with the environment without an explicit model of the system dynamics. In our MECCO scenario, at the beginning, the agent has no experience and information about the MECCO environment. Thus it needs to \textit{explore} for every time epoch by taking some actions at each offloading state, e.g., the size of current IoT data size, available edge resource. As long as the agent has some experiences from actual interactions with the environment, it will \textit{exploit} the known information of states while keep exploration. As a combination of the Monte Carlo method and dynamic programing, a temporal-difference (TD) approach can be employed to allow the agent to learn offloading policies without requiring the state transition probability which is difficult to acquire in realistic scenarios like in our dynamic mobile blockchain. Therefore, we can develop a dynamic offloading scheme using a free-model RL. Specially, in this paper, our focus is to find the optimal policy that minimize the offloading cost $C$. To this end, the state-action function can be updated using the experience tuple of agent $(s^t,a^t, r^t, s^{t+1})$ at each time step t in our offloading application as
\begin{multline}
Q(s^t,a^t) \leftarrow Q(s^t,a^t) + \alpha[r(s^t,a^t) + \gamma*minQ(s^{t+1},a^{t+1}),\\
-  Q(s^t,a^t) ]
\end{multline}
which is called as Q-learning algorithm \cite{14}. Here $\alpha$ is the learning rate, $\gamma$ is the discount factor between (0,1) and $\sigma^t = r(s^t,a^t) + \gamma*maxQ(s^{t+1},a^{t+1}) - Q(s^t,a^t)$ is the TD error which will be zero for the optimal Q-value. Further, under the optimal policy $\pi^*$ which can be obtained from the maximum Q-value ($\pi^*(s) = arg maxQ^*(s,a)$), the Bellman optimality equation \cite{16} for the state-action equation can be expressed as 
\begin{equation}
Q^*(s^t,a^t) = \mathds{E}_{s^{t+1} \sim E}[ r(s^t,a^t) + \gamma*minQ^*(s^{t+1},a^{t+1})].
\end{equation}

It is noting that the Q-learning algorithm is proved to converge with probability one over an infinite number of times \cite{25}  and achieves the optimal $Q^*$. Although the reinforcement learning can solve the offloading problem by obtaining the optimum reward, there are still some remaining problems. The state and action values in the Q-learning method are stored in a two-dimensional Q table, but this method can become infeasible to solve complex problems with a much larger state-action space. This is because if we keep all Q-values in a table, the matrix $Q(s,a)$ can be very large, which makes the learning agents difficult to obtain sufficient samples to explore each state, leading to the failure of the learning algorithm. Moreover, the algorithm will converge slowly due to too many states that the agent has to process. 

To overcome such challenges, we can use deep learning with Deep Neural Network (DNN) to approximate the Q-values instead of using the conventional Q-table, leading to a new algorithm called DRL. In the DRL-based algorithm, a DNN is used to approximate the target Q-values $Q(s^t,a,\theta) $ with weights $\theta$. Further, to solve the instability of Q-network due to function approximation, the experience replay solution is employed in the training phase with the buffer $\mathcal{B}$ which stores experiences $e^t= (s^t,a^t,r^t,s^{t+1})$ at each time step $t$. Next, a random mini-batch of transitions $(s^j,a^j,r^j,s^{j+1})$ from the replay memory is selected to train the Q-network. Here the Q-network is trained by iteratively updating the weights $\theta$ to minimize the loss function, which is written as
\begin{equation}
L(\theta) = \mathds{E}  [y^j-Q(s^j,a^j|\theta^j))^2],
\end{equation}
where $y^j=(r^j +\gamma*min_{a^{j+1}}Q(s^{j+1},a^{j+1}|\theta')$ and the $\mathds{E}[.]$  denotes the expectation function.

\subsubsection{Advanced DRL-based Computation Offloading}  
Recent years have witnessed great efforts in deep reinforcement learning to improve the performance of DLR-based algorithm. In this paper, we focus on two recent improvements in DRL research to apply to our formulated offloading problem, including double DQN  and dueling DQN \cite{24}. 
\begin{itemize}
	\item \textit{Double DQN:} In the conventional DQN, we use the same samples from the replay memory to both specify which action is the best and estimate this action value, which leads to the large over-estimation of action values. To solve this problem, two Q-functions are proposed to select and evaluate the action values by a new loss function as:
	\begin{equation}
	L_{dou}(\theta) = \mathds{E}  [y^j_{dou}-Q(s^j,a^j|\theta^j))^2],
	\end{equation}
	where
	\begin{equation}
	y^j_{dou} = (r^j+\gamma.Q(s^{j+1},min_{a^{j+1}}Q(s^{j+1},a^{j+1}|\theta^1),\theta^2).
	\end{equation}
	It is noting that the action choice is still based on the weight $\theta^1$, while the evaluation of the selected action relies on the weight value $\theta^2$. This technique mitigates over-estimation problem and thus improves the training process and in turn the efficiency of the DQN algorithms. 
	\item \textit{Dueling DQN: } During the computation of Q function in some MDP problems, it is unnecessary to estimate action and state values at the same time, thus we can estimate separately the action and state value functions. Motivated by this concept, in dueling DQN, the state action value $Q(s,a)$ can be decomposed into two value functions as follows:
	\begin{equation}
	Q(s,a) = V(s) + A(a).
	\end{equation}
	Here, $V(s)$ is the state-value function which evaluates the significance of being at a given state $s$. $A(a)$ is action-value function to estimate the significance of choosing an action $a$ compared to other actions. The $V(s)$ and $A(a)$ are first computed separately, then are combined to generate the final output $Q(s,a)$. This approach leads to a better performance on policy evaluation of MDPs, especially complex problems with large action space like our considered offloading scenario.
\end{itemize}
Motivated by the advantages of the above approaches, we develop a novel DQN algorithm using such two improvements to solve the offloading problem of our MECCO system. The details of the proposed algorithm is shown in Algorithm~\ref{Alg:Offloading}. The ADRLO algorithm can achieve the optimal task offloading strategy in an iterative manner. Here, the procedure generates a task offloading strategy for MDs based on system states and observes the system reward at each time epoch so that the offloading policy can be optimized (lines 8-16). Then the procedure updates the history experience tuple and train the Q-network (lines 18-22) with loss function minimization. This trial and error solution will avoid the requirement of prior information of offloading environment. Over the training time period, the trained deep neural network can characterize well the environment and therefore, the proposed offloading algorithm can dynamically adapt to the real MECCO environment.

\begin{algorithm}
	\footnotesize
	\caption{Advanced DRL-based computation offloading (ADRLO) algorithm for MECCO system}
	\begin{algorithmic}[1]
		\STATE \textbf{Initialization:}
		\STATE Set replay memory $\mathcal{D}$ with capacity $N$
		\STATE Initialize the deep Q network  $Q(s,a)$ with random weight $\theta$ and $\theta^'$, initialize the exploration probability $\epsilon \in (0,1)$
		\FOR{episode = 1,..., \textit{M}}
		\STATE Initialize the state sequence $s^0$
		\FOR{$t = 1,2,...$}
		\STATE \textit{/$***$ Plan the computation offloading $***$/}
		\STATE Estimate the current offloading cost $tc^t$
		\STATE Estimate the available edge resource $ec^t$
		\STATE Estimate the available bandwidth resource $bw^t$
		\STATE Estimate the consumed gas cost $g^t$
		\STATE Set $s^t = \{tc^t, ec^t, bw^t, g^t \}$
		\STATE Select a random action $a^t$ with probability $\epsilon$, otherwise $a^t = argminQ(s^t,a,\theta)$
		\STATE Offload the computation task $\alpha_e^t(D^t)$ to the edge or cloud server $\alpha_c^t(D^t)$  
		\STATE Observe the reward $r^t$  and  next state $s^{t+1}$
		\STATE Evaluate the system cost $C(s,a)^t$
		\STATE \textit{/$***$ Update $***$/}
		\STATE Store the experience ($s^t,a^t,r^t,s^{t+1}$) into the memory $\mathcal{D}$
		\STATE Sample random mini-batch of state transitions ($s^j,a^j,r^j,s^{j+1}$) from  $\mathcal{D}$
		\STATE Calculate the target Q-value by ($y^j_{dou} = r^j+\gamma.Q(s^{j+1},min_{a^{j+1}}Q(s^{j+1},a^{j+1}|\theta),\theta^')$
		\STATE Perform a gradient descent step with the weight $\theta^j$ on $ (y^j_{dou}-Q(s^j,a^j|\theta^j))^2$
		as the loss function
		\STATE Train the deep Q-network with updated $\theta$ and $\theta^'$
		\ENDFOR 
			\ENDFOR 
	\end{algorithmic}
	\label{Alg:Offloading}
\end{algorithm}

\begin{algorithm}
	\caption{Joint access control and computation offloading on blockchain for IoT networks}
	\begin{algorithmic}[1]
		\STATE Initialize private blockchain, and setup \textit{N} Ethereum nodes for all MDs
		\STATE Deploy smart contracts on the cloud
		\STATE Blockchain starts mining
		\STATE \textbf{Access control phase:}
		\FOR{each MD \textit{n} in \textit{N}}
			\STATE Initialize the transaction \textit{Tx} for offloading requests
			\STATE Submit the transaction \textit{Tx} to MECCO manager
			\STATE Perform access control via Algorithm~\ref{Alg:AccessCOntrol}
			\IF{verification is successful}
				\STATE Go to the offloading phase
			\ELSE
				\STATE Reject the offloading request 
			\ENDIF
		\ENDFOR
		\STATE \textbf{The offloading phase:}
		\FOR {each authorized MD \textit{n} in $N^'$}
			\STATE Perform task offloading via Algorithm~\ref{Alg:Offloading}
			\STATE Verify the offloading transaction \textit{Tx} by mining process
			\STATE Upload the transaction \textit{Tx} to blockchain and wait for confirmation
		\ENDFOR	
	\end{algorithmic}
\label{Alg:TotalAlg}
\end{algorithm}

To this end, we propose a novel MECCO algorithm by combining access control and computation offloading on a mobile blockchain IoT network. The concept of the integrated scheme is shown in Algorithm~\ref{Alg:TotalAlg}, which can be explained as follows. We first create a private Ethereum blockchain environment on cloud, i.e., Amazon cloud platform, to perform access control and offloading functionalities. With blockchain setup, we can deploy a smart contract and connect with a network of MDs, e.g., smartphones, to formulate a MECCO system (lines 1-3). It is assumed that all MDs have the demand to offload their computation tasks to edge or cloud servers for execution. In the access control phase, each MD will send an offloading request as a transaction in blockchain context to the cloud server for authentication. The smart contract deployed on cloud will verify the transaction to accept or refuse the request. If the request is accepted, the process now goes to the offloading phase, otherwise a penalty will be given to the requestor (lines 4-14). In the offloading phase, it is assumed that all authorized MDs in the access control phase are grouped into the new set of devices ($N^'$). We perform the offloading algorithm using a DRL network to optimize the offloading cost for our MECCO. The process finishes with the mining of the transaction and appending it to the blockchain (lines 15-20). 

\textcolor{black}{The computational complexity of our algorithm mostly comes from the complexity of running the DNN. We assume that are $K$ neurons at the input layer of the DNN, and $Z$ as the number of neurons at the output layer. Also, the hidden layer is $L$, and the number of neurons at hidden layers is $H$. Accordingly, the computation cost at the DNN is $(KH+(L-1)HH+HZ)$ =$ \mathcal{O}(H(K+(L-1)H+Z))$. Also, the complexity of using activation function is $ \mathcal{O}(HL)$. Hence, the total complexity is $ \mathcal{O}(H(K+HL-H+Z+L))$ which can be simplified as $ \mathcal{O}(H(K+HL+Z))$.}

\section{Simulations and Performance Evaluations}
\label{Sec:Evaluation}
In this section, we investigate the proposed MECCO system by conducting both real experiments to evaluate the access control performance and numerical simulations to evaluate the efficiency of our proposed secure computation offloading scheme.
\subsection{Implementation Settings}

\begin{figure}
	\centering
	\includegraphics[width=0.5\linewidth]{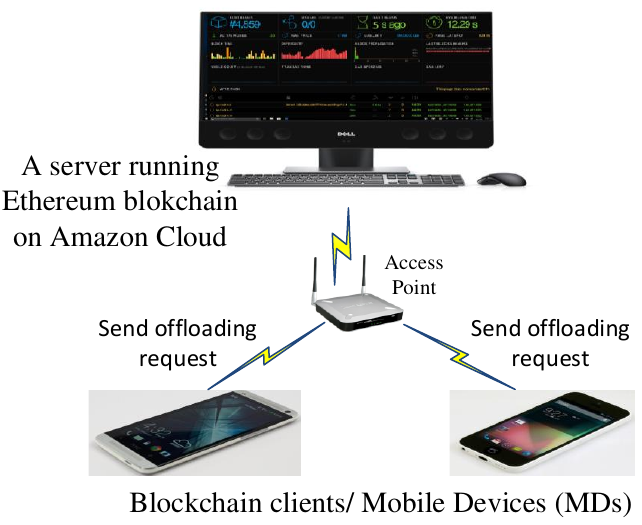}
	\caption{Experiment setup for access control implementation.}
	\label{Fig:Expere}
	\vspace{-0.16in}
\end{figure}

We considered an access control framework for MECCO on a mobile cloud as shown in Fig.~\ref{Fig:Expere}. \textcolor{black}{We deployed a private Ethereum blockchain which is available on the Amazon cloud\footnote{https://aws.amazon.com/blockchain/} computing platform}, where two virtual machines AWS EC2 were employed as the miners, two virtual machines Ubuntu 16.04 LTS were used as the admin and MECCO manager, respectively. Our smart contract was written by Solidity programming language \cite{30} as shown in Fig.~\ref{Fig:CodeSC} and was deployed on AWS Lambda functions. Each function interacts with the cloud blockchain via the web3.js API. MDs can interact with smart contracts through their Android phone where a Geth client (a command line interface implemented in the Go language) was installed to transform each smartphone into an Ethereum node. By using the Geth client, a MD can create an Ethereum account to communicate with our blockchain network for accessing data. The web3.js library, a lightweight Java library for working with smart contracts and blockchain, was also used for developing the mobile application to connect with the Ethereum blockchain network. In our experiment in this paper, we used two Sony mobile phones running on an Android OS version 8.0 platform to investigate the access control results. 

Moreover, in our simulation for computation offloading, a MECCO system is considered with a cloud server, a MEC server with a number of MDs authorized by the access control mechanism. Here we consider $N=[2-30]$ MDs, each of them has a computation task to be executed at edge or cloud server. We assume that the data size of IoT computation tasks $D$ is randomly distributed between 0.1MB and 12MB \cite{30}. The transmit power $p_n$ is 0.4 W. The total bandwidth resource $B$ is set to 15 MHz \cite{25} ; the additive noisy power spectral density $N_0$ is -100dBm/Hz \cite{24,25}. Besides, the total computational  capacity of MEC server $F^e$ and cloud server $F^c$ are set to 2GHz and 5GHz, respectively \cite{23}. The required number of CPU cyles $X_n$ is set to $[0.8-1.5]$ Gcyles. The completion deadline for executing a task $tau_n$ is 1000 ms. The channel gain between MDs and the MEC server $h_n$ is generated using distance-dependent path-loss model $L[dB] =140.7 + 36.7log_{10}d_{[km]}$ \cite{25,30}. The factors $\beta_n^t, \beta_n^e$ are set to 0.6 and 0.4, respectively. Also, the gas price $\xi$ at the cloud is set to $2e^{-9}$ and the contract gas demands are $g_n$ are $[10^4-10^9]$ gas \cite{13}. 

\textcolor{black}{Regarding the DRL algorithm, the architecture of the DNN-based training model needs to be built carefully. The increase of the number of hidden layers will make the algorithm more complex and the training takes much time. However, if the numbers of hidden layers and neurons on each layer are too small, the algorithm may not achieve a desired convergence. In this paper, the parameters are selected by performing multiple tests and numerical simulations. In this particular work, the discount factor $\gamma$ is set to 0.85 and the replay memory capacity and training batch size are 105 and 128, respectively. The used DNN structure includes three hidden layers, where the first, second, and third hidden layers have 64, 32 and 32 neurons, respectively. We employ ReLU as the activation function in the hidden layers, while the sigmoid activation function is utilized in the output layer to relax the offloading decision variables.} The simulations were implemented in Python with TensorFlow 2.0 on a computer with an Intel Core i7 4.7GHz CPU and 128 GB memory.

\textcolor{black}{The concept of data training in our DRL algorithm can be explained as follows. Each episode consists of 50 time slots (or steps), and the DRL training is performed via 2000 episodes. At the beginning of each time step in a training episode, the action set is chosen randomly, including offloading decision, edge resource, and bandwidth resource values. Then, the agents (MDs) interact with the offloading environment (established by MDs, MEC and cloud servers) to compute the computation latency, energy consumption, and the smart contract fee values of the offloading process based on the predefined functions described in the previous sections. This aims to create a matrix of data for computing the reward later. Then, the current state and action are stored, and the next state and the reward value are updated ready for the next step of training. The training is iterated within the given episode until the algorithm converges. }
\subsection{Access Control Performance}
\begin{figure}
	\centering
	\includegraphics[width=0.98\linewidth]{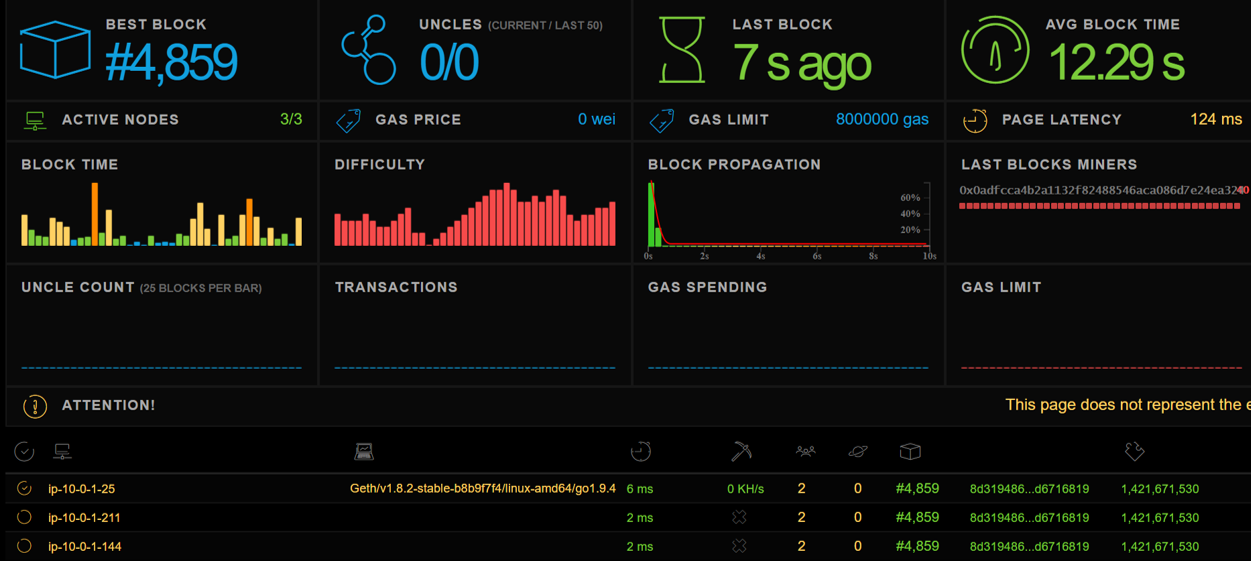}
	\caption{The running Ethereum blockchain on Amazon cloud for access control of MECCO system.}
	\label{Fig:AWS}
	\vspace{-0.15in}
\end{figure}

In this subsection, we implemented access control on blockchain and verified the proposed scheme. We first deployed a private Ethereum blockchain on Amazon cloud as illustrated in Fig.~\ref{Fig:AWS}. Offloading access and transactions are recorded and shown on the web interface for monitoring. Based on our blockchain settings, we deployed smart contracts, established network entities of the access control as explained in Fig.~\ref{Fig:AccessControl} and connected with mobile applications to build our access control framework. 

\begin{figure*}
	\centering
	\includegraphics[width=0.98\linewidth]{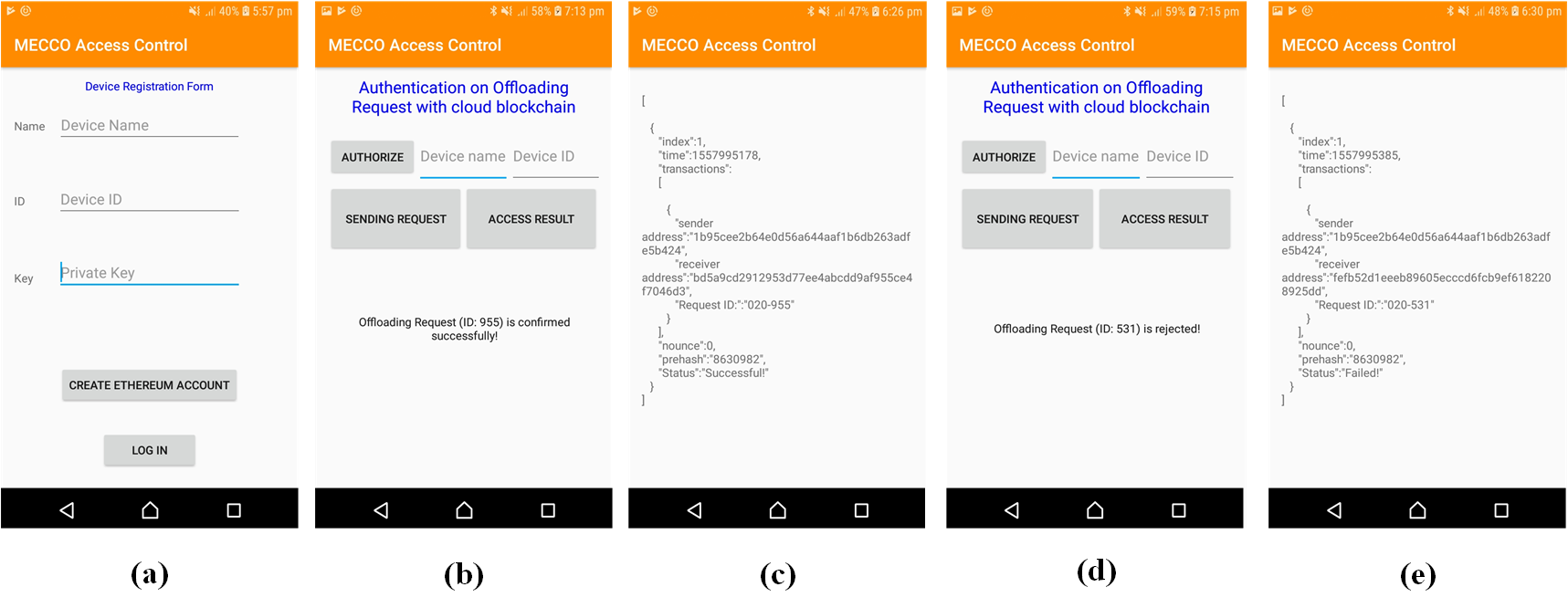}
	\caption{Illustrations of access control results.}
	\label{Fig:Simulation_AccessControlResult}
	\vspace{-0.15in}
\end{figure*}
We evaluated the performance of the proposed access control scheme via two use cases with authorized and unauthorized access for offloading request, as illustrated in Fig.~\ref{Fig:Simulation_AccessControlResult}.  The main objective of the access control is to verify effectively the offloading request from the MDs and prevent any potential threats to our MECCO system on mobile edge-cloud. The evaluation is presented as follows. First, it is assumed that an owner of a smartphone  wants to offload its data task to the cloud or edge server for calculation. He should create an Ethereum blockchain account and register an access right of his device by providing the device name, device ID and make a transaction for an offloading request (Fig.~\ref{Fig:Simulation_AccessControlResult}(a)).  Note that based on the blockchain concept, the transaction is signed by a private key coupled with a public key for device identification. After receiving the transaction from the MD, the MECCO manager will authorize the request using the smart contract. If the MD is verified by the smart contract, the request is now confirmed for task offloading. Accordingly, a response is given to the requestor for confirmation so that the MD can offload its data to edge or cloud server (Fig.~\ref{Fig:Simulation_AccessControlResult}(b)). Now the access control process finishes and the offloading transaction is appended to blockchain by the cloud miner and broadcast to all entities in the network. Therefore, a MD can keep track of the offloading access (Fig.~\ref{Fig:Simulation_AccessControlResult}(c)), which improves network trustworthiness.

In the case of unauthorized access, the smart contract will verify and detect by the access protocol with a predefined policy list. Such illegal request is prevented and discarded from blockchain, and a warning message is returned to the requester (Fig.~\ref{Fig:Simulation_AccessControlResult}(d)). A corresponding transaction for unauthorized access is also issued by the smart contract (Fig.~\ref{Fig:Simulation_AccessControlResult}(e)). Obviously, the use of blockchain can address effectively challenges mentioned in the literature in controlling access information and monitoring offloading behaviours, which can enhance system reliability and data privacy.

\subsection{Computation Offloading Performance}
\begin{figure}[t!]
		\centering
		\includegraphics[width=0.99\linewidth]{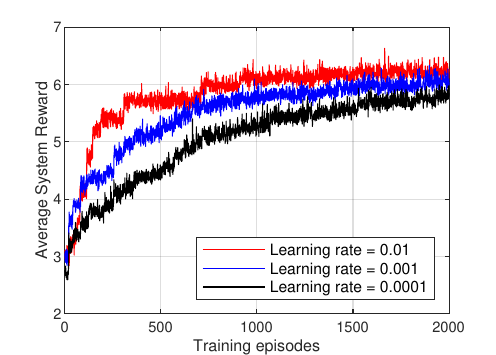} 
		\caption{Comparison of average system rewards with different learning rates. }
		\label{Fig:Learning}
\end{figure}

\begin{figure*}
	\centering
	\begin{adjustbox}{max width=1\textwidth}
		\subfigure[Total cost versus the number of MDs.]{
			\begin{tikzpicture}
			\begin{axis}[legend pos=north west, legend cell align={left}, 
			xlabel=Number of Mobile Device,grid=major,ymax=55,
			ylabel=Total Cost,  every axis plot/.append style={ultra thick}]
			\addplot coordinates {
				(2,7.85)
				(4, 8.95)
				(6,10.3)
				(8,11.5)
				(10,12.25)
				(12,13.05)	
			};
			\addplot coordinates {
				(2,9.95)
				(4,11.2)
				(6,12.6)
				(8,14)
				(10,15.25)
				(12,16.35)	
			};
			\addplot [mark =triangle*, brown] coordinates {
				(2,11.85)
				(4,15.7)
				(6,18.55)
				(8,21.9)
				(10,24.85)
				(12,27.3)
			};
			\addplot coordinates {
				(2,13.05)
				(4,17.6)
				(6,22.6)
				(8,26.6)
				(10,30.6)
				(12,34.1)
			};
			\legend{The proposed ADRLO scheme, The DRL scheme \cite{14}, The EO scheme \cite{10}, The CO scheme \cite{3}}
			\end{axis}
			\end{tikzpicture}	}
		\subfigure[Total cost versus the number of tasks.]{
			\begin{tikzpicture}
			\begin{axis}[legend pos=north west, legend cell align={left}, 
			xlabel=Task Size (MB),grid=major,ymax=125,
			ylabel=Total Cost,  every axis plot/.append style={ultra thick}]
			\addplot coordinates {
				(2,31)
				(4, 36)
				(6,41)
				(8,45)
				(10,49)
				(12,52)			
			};
			\addplot coordinates {
				(2,36)
				(4,41)
				(6,46)
				(8,51)
				(10,55)
				(12,59)				
			};
			\addplot [mark =triangle*, brown] coordinates {
				(2,42)
				(4,46)
				(6,54)
				(8,67)
				(10,81)
				(12,95)
				
			};
			\addplot coordinates {
				(2,58)
				(4,52)
				(6,50)
				(8,55)
				(10,62)
				(12,72)
			};
			\legend{The proposed ADRLO scheme, The DRL scheme \cite{14}, The EO scheme \cite{10}, The CO scheme \cite{3}}	
			\end{axis}
			\end{tikzpicture}	}
	\end{adjustbox}
	
	\caption{Total offloading cost versus the number of MDs and tasks.} \label{Fig:OffloadingCost_MDs}
\end{figure*}
We first evaluate the training performance of our ADRLO algorithm. Fig.~\ref{Fig:Learning} shows the performance of average system reward with different learning rates. It can be seen that the learning rate affects the learning rewards over the training episodes. That is, when the learning rate decreases, the convergence performance of the proposed algorithm decreases due to slow learning speed. Based on our experimental results, the learning value at  0.01 yields the best reward performance and has good convergence rate and thus we use it in the following system simulations and evaluations.

We then evaluate the proposed ADRLO scheme  in various performance metrics with extensive analysis on offloading costs of latency and energy consumption under different parameter settings. To highlight the advantage of the proposed offloading scheme in terms of offloading cost efficiency, we compare our ADRLO algorithm with the following baseline schemes, i.e., 
\begin{itemize}
	\item DRL-based offloading scheme (DRLO) \cite{14}: Computation offloading of MECCO system is performed by an offloading algorithm using regular DRL. 
	
	\item Edge offloading scheme (EO) \cite{10}:  All MDs offload their computation task to the MEC server, .i.e, setting offloading decision vector $( \alpha_n^e=1, \alpha_n^c=0$, $(\forall \textit{n} \in \mathcal{N}))$.
	
	\item Cloud offloading scheme (CO) \cite{3}:  All MDs offload their computation task to the cloud server, .i.e, setting offloading decision vector $( \alpha_n^c=1, \alpha_n^e=0$, $(\forall \textit{n} \in \mathcal{N}))$.
\end{itemize}

Further, to investigate the performance of different offloading schemes, the simulation results are averaged over 50 runs of numerical simulations.

\begin{figure*}
	\centering
	\begin{adjustbox}{max width=1\textwidth}
		\subfigure[Total cost versus edge resource.]{
			\begin{tikzpicture}
			\begin{axis}[legend pos=north east, legend cell align={left}, 
			xlabel=Computational Capacity of MEC server (GHz),grid=major,ymax=110,
			ylabel=Total Cost,  every axis plot/.append style={ultra thick}]
			\addplot coordinates {
				(0.5, 16	)
				(1, 	14)
				(1.5,	11)
				(2,		10)
				(2.5, 	9.5)
				(3,		8.7)	
				(3.5,	8.2)
				(4, 	7.4)
				(4.5,	6.9)
				(5,		6.0)

			};
			\addplot coordinates {
						(0.5, 21	)
						(1, 	17)
						(1.5,	14)
						(2,		13)
						(2.5, 	12.5)
						(3,		11.7)	
						(3.5,	10.6)
						(4, 	9.4)
						(4.5,	8.9)
						(5,		8)
		};
			\addplot [mark =triangle*, brown] coordinates {
				(0.5, 	88)
				(1, 	71)
				(1.5,	60)
				(2,		52)
				(2.5, 	46)
				(3,		41)	
				(3.5,	37)
				(4, 	33)
				(4.5,	29)
				(5,		26)
			};
			\addplot coordinates {

				(0.5, 28)
				(1, 	28)
				(1.5,	28)
				(2,		28)
				(2.5, 	28)
				(3,		28)	
				(3.5,	28)
				(4, 	28)
				(4.5,	28)
				(5,		28)
			};
			\legend{The proposed ADRLO scheme, The DRL scheme \cite{14}, The EO scheme \cite{10}, The CO scheme \cite{3}}
			\end{axis}
			\end{tikzpicture}	}
		\subfigure[Total cost versus bandwidth resource.]{
			\begin{tikzpicture}
			\begin{axis}[legend pos=north east, legend cell align={left},
			xlabel=Total System Bandwidth (MHz),grid=major,ymax=75,
			ylabel=Total Cost,  every axis plot/.append style={ultra thick}]
			\addplot coordinates {
				(1, 19)
				(2, 15.5)
				(3, 13.)
				(4, 11)
				(5, 9.5)
				(6, 8.4)	
				(7, 7.2)
				(8, 6)
				(9, 5.3)
				(10, 5)		
			};
			\addplot coordinates {
				(1, 21)
				(2, 17)
				(3, 14)
				(4, 12)
				(5, 10.5)
				(6, 9)	
				(7, 7.5)
				(8, 6.3)
				(9, 5.6)
				(10, 5.3)				
			};
			\addplot [mark =triangle*, brown] coordinates {
				(1, 36)
				(2, 32.5)
				(3, 29)
				(4, 27)
				(5, 25)
				(6, 23.5)	
				(7, 22)
				(8, 20.6)
				(9, 19)
				(10, 17.6)	
			};
			\addplot coordinates {
				(1, 48)
				(2, 41)
				(3, 37)
				(4, 34.4)
				(5, 32.2)
				(6, 30.5)	
				(7, 28.4)
				(8, 26.7)
				(9, 25)
				(10, 24)	
			};
			\legend{The proposed ADRLO scheme, The DRL scheme \cite{14}, The EO scheme \cite{10}, The CO scheme \cite{3}}		
			\end{axis}
			\end{tikzpicture}	}
	\end{adjustbox}
	
	\caption{Total offloading cost versus resource allocation.} \label{Fig:OffloadingCost_Resource}
	\vspace{-0.15in}
\end{figure*}

\begin{figure}
	\centering
	\resizebox {8.9cm} {7.2cm} {
		\begin{tikzpicture}
		\begin{axis}[legend pos=north west, legend cell align={left}, 
		xlabel=Task Size (MB),grid=major, style={font=\small},
		ylabel=Total Cost,ymax= 65,  every axis plot/.append style={ultra thick}]
		\addplot coordinates {
			(2, 	28)
			(4, 	31.5)
			(6,		35.5)
			(8,		39.5)
			(10, 	43.5)
			(12,	47.5)	
		};
		\addplot coordinates {
			(2, 	33)
			(4, 	36.5)
			(6,		40.5)
			(8,		44.5)
			(10, 	48.5)
			(12,	52.5)
		};
		\addplot [mark =triangle*, brown] coordinates {
			(2, 	31)
			(4, 	34.5)
			(6,		38.5)
			(8,		42.5)
			(10, 	46.5)
			(12,	50.5)
		};
		\legend{Proposed ADRLO scheme,ADRLO scheme w/o edge allocation, ADRLO scheme w/o bandwidth allocation}	
		\end{axis}
		\end{tikzpicture}	}
	\caption{Total costs of offloading schemes.} 
	\label{Fig:OffloadingCost_TaskSize}
\end{figure}

\begin{figure}
	\centering
	\resizebox {8.9cm} {7.2cm} {
		\begin{tikzpicture}
		\begin{axis}[legend pos=north west, legend cell align={left}, grid=major,
		xlabel=Number of Mobile Device,
		ylabel=The Smart Contract Cost (Ether) ,  every axis plot/.append style={ultra thick}]
		\addplot coordinates {
			(5,0.092)
			(10,0.096)
			(15,0.102)
			(20,0.112)
			(25,0.130)
			(30,0.146)	
			
		};
		\addplot coordinates {
			(5,0.094)
			(10,0.100)
			(15,0.107)
			(20,0.121)
			(25,0.140)
			(30,0.159)
			
		};
		\legend{The proposed ADRLO scheme, The DRL scheme \cite{14}}
		\end{axis}
		\end{tikzpicture}	
	}
	\caption{Smart contract cost for offloading authentication versus the number of MDs.} 
	\label{Fig:SmartConstract_Cost}
\end{figure}
We first evaluate the performance of the total MECCO cost versus the number of MDs and tasks in Fig.~\ref{Fig:OffloadingCost_MDs}. More specifically, in Fig.~\ref{Fig:OffloadingCost_MDs}(a), we consider a MECCO system with a varying number of MDs, and each MD has a computation task whose size varies between $0.1$ MB and $1$ MB. As indicated from the simulation results, the curves of all offloading schemes increase with the growing number of MDs. Specially, the CO scheme has the highest offloading cost among all offloading schemes. An explanation is that the computation tasks of MDs in this test case are relatively small, so offloading to the remote cloud will incur a larger transmission latency, and thus leads to a higher offloading cost. Meanwhile, thanks to low-latency computation services of MEC server, the EO scheme shows better offloading efficiency with lower offloading costs for any user cases. More importantly, the DRLO and ADRLO schemes exhibit much lower offloading costs and the ADRLO scheme achieves the best performance among all offloading schemes, which can be explained by the following reasons. First, in the proposed advanced DRL algorithm, the double DQN network obtains the optimal policy with larger system gains than that of the regular DQN.  Second, the dueling DQN architecture leads to a better performance on policy evaluation by evaluating separately the state-value function and the action-value function. These advanced techniques make the ADRLO scheme more efficient in terms of evaluation of optimal offloading policy and offloading performance. 

Next, we analyze the computation offloading performance for the MECCO system with a single MD \textit{N} = 1 and its task size changes between 2MB and 12MB as shown in Fig.~\ref{Fig:OffloadingCost_MDs}(b). It is observed that when the task size increases, the cost of four schemes increases due to the growing amount of IoT data to be executed completely. Particularly, when the task size is small ($<$5MB), the CO scheme has a higher offloading cost than that the EO scheme. The reason behind this observation is the small tasks can be processed efficiently by the MEC server with sufficient computation resources. Therefore, offloading small tasks to the remote cloud will result in unnecessary transmission latency and consequently, incur a higher total offloading cost for the CO scheme. However, when the task size become larger ($>$5MB), the computational capacity of the MEC server becomes less sufficient to accommodate all tasks, while the resourceful cloud server can compute large-size tasks effectively. As a result, the CO scheme can achieve a much lower offloading cost, compared to the EO scheme. Note that the ADRLO scheme still achieves the minimum total offloading cost, followed by the DRLO scheme with a small gap when the task size increases.  

Moreover, we compare offloading schemes under resource allocation scenarios in Fig.~\ref{Fig:OffloadingCost_Resource}. First, we evaluate the impact of edge resource allocation on offloading performance.  From Fig.~\ref{Fig:OffloadingCost_Resource}(a), the total offloading cost of schemes based on edge computing decreases significantly with the increase of the computational capacity of MEC server. Furthermore, the CO scheme has a stable offloading cost with the increase of MEC capacity as all computation tasks in this scheme are processed in the cloud. In particular, when the edge computational capacity is small, the EO scheme has the highest offloading cost. This is because a less computing resource of MEC server will lead to a higher execution latency, and thus the EO scheme suffers from a much higher computation cost, compared to other schemes. However, when the edge computational capacity becomes large, the total system cost of the EO scheme will reduce significantly and can achieve a lower cost than that the CO scheme (i.e., when the edge capacity is 5GHz). This result can provide more insights into how to allocate properly allocation for MEC server to enhance the overall offloading performance for the MECCO system.  Again, the proposed ADRLO scheme can obtain the lowest cost with a downward trend and thus exhibits the best performance when compared to the DRLO scheme and other baselines. 

Meanwhile, Fig.~\ref{Fig:OffloadingCost_Resource}(b) shows the total offloading cost versus the system bandwidth allocation. Based on the simulation result, it is clear that all the curves of offloading schemes decrease gradually with the increment of the total system bandwidth. The reason behind this observation that bandwidth allocation is always significant in improving the data transmission rate between MDs and edge-cloud server. As a result, this solution will reduce the transmission delay and the total offloading cost, accordingly. Further, by applying the proposed dynamic offloading policy, our approach using reinforcement learning can adjust dynamically how much bandwidth resource should be allocated to a certain MD based on MD’s task size so as to achieve the optimal offloading for the MECCO system. 

We also analyze the efficiency of the proposed ADRLO scheme by comparing to the other two baseline solutions: the proposed scheme without edge allocation and the proposed scheme without bandwidth allocation. Note that for these baselines, edge computation and bandwidth resources are allocated equally to all MDs in the MECCO system regardless of the task sizes. As indicated in Fig.~\ref{Fig:OffloadingCost_TaskSize}, the proposed scheme, which takes edge resource and bandwidth resource into account, can achieve the best performance in terms of minimum total offloading costs, compared to other benchmarks. Obviously, by jointly optimizing both offloading decision and resource allocation, our proposed algorithm can achieve effective computation cost savings and improve significantly the offloading performance of the MECCO system.

We also evaluate the smart contract cost for offloading authentication, as illustrated in Fig.~\ref{Fig:SmartConstract_Cost}, under our proposed ADRLO scheme and the baseline DRLO scheme. It is indicated that more MDs lead to the higher contract cost because more offloading transactions need to be verified and executed by the smart contract which requires more gas costs to achieve offloading on edge cloud. However, the contract cost achieved by the proposed ADRLO scheme is lower than the  DRLO approach with different numbers of MDs, thanks to the optimized learning policy. 

{\color{black}Finally, we investigate the execution latency of our DRL algorithm. For fair comparison, we use a RL approach and also consider a popular exhaustive search approach, where the MEC server collects all global information over the large state-action space and computes the offloading costs for all MDs in each offloading realization. As shown in Table~\ref{table:latency}, our DRL approach yields lower latency than the RL approach, and it has a much lower running time for each offloading realization, compared to the greedy approach. For example, when the number of MDs is 60, our DRL algorithm generates an offloading decision in about 5.891 second for each realization, while the RL and greedy approach spends 5 times and 14 times longer execution time, respectively. This is because that the RL algorithm with table searching-based Q-learning method cannot achieve the optimal offloading policy due to the problem of high curse-of-dimensionality when we optimize jointly multiple edge and bandwidth resources concurrently. Moreover, the exhaustive search approach need much time to find the optimal offloading action by the evaluation over the large state-action space. Overall, our DRL approach with a double-dueling Q network take advantage of the DNN to train the Q-learning process for more efficient offloading optimization. } Finally, the advantages of our scheme over existing approaches are summarized in Table~\ref{table:FeatureComparisons}. 

\begin{table}
	\scriptsize
	\centering
	\captionsetup{font=scriptsize}
	\caption{{\color{black}Comparison of average execution latency for each offloading realization (in second).}}
	\label{table:latency}
	{\color{black}
	\begin{tabular}{|c||c c c|}
		\hline
		\textbf{No. of MDs}  & \textbf{Our DRL approach} & \textbf{RL approach} & \textbf{Exhaustive Search} \\
		\hline
		N= 20 & 2.241&	6.704&	27.034 \\
		
		N= 40 &	3.590&	10.501&	40.440\\
		N =60 &	5.891&	16.093&	82.063 \\
		\hline
	\end{tabular}}
\end{table}

\begin{table}
	\scriptsize
	\centering
	\captionsetup{font=scriptsize}
	\caption{{\textcolor{black}{Comparison between our proposed scheme and the existing works.}}}
	\textcolor{black}{\begin{tabular}{|p{2cm}||c |c| c| c| c| c|c|}
			\hline
			\multirow{2}{*}{\textbf{Features}} &
			\multicolumn{7}{c|}{\textbf{Schemes}} \\
			&\cite{12}&	\cite{15}&	\cite{24}&	\cite{25}  &	\cite{17}&	\cite{19}&	Ours \\
			\hline
			Intelligent computation offloading&	\checkmark&	\checkmark&	\checkmark&	\checkmark&	&	&	\checkmark
			\\ \hline
			Advanced DRL design with double-dueling Q learning&	&	&	\checkmark&	&	&	&	\checkmark
			\\ \hline
			Smart contract design&	&	&	&	&	\checkmark&	\checkmark&	\checkmark
			\\ \hline
			Offloading design in blockchain&	\checkmark&	\checkmark&	&	\checkmark&	&	&	\checkmark
			\\ \hline
			Smart contract-based security and offloading services&	&	&	&	&	&	&	\checkmark
			\\ \hline
			%		Real-world experiments for mobile blockchain-based IoT &	&	&	&	&\checkmark	&\checkmark	&	\checkmark
			%		\\ \hline
	\end{tabular}}
	\label{table:FeatureComparisons}
	\vspace{-0.1in}
\end{table}

\section{Conclusions}
\label{Sec:Conclude}
In this paper, we have jointly studied access control and computation offloading for blockchain-based computation offloading, which has not been considered fully in the literature. First, to improve  the security of computation task offloading, we propose a new access control mechanism enabled by smart contracts and blockchain to manage access of MDs with the objective of preventing malicious offloading access and preserving cloud resources. Then, we propose a novel DRL-based offloading scheme to obtain the optimal offloading policy for all MDs in the IoT network. We formulate task offloading decision, edge resource allocation, bandwidth allocation, and smart contract usage as a joint optimization problem, which is solved efficiently by an advanced DQN algorithm with a double-dueling Q-network in a fashion that the total offloading cost of computation latency, energy consumption and smart contract cost are minimized.  The implementation results on both real-world exeperiments and simulations showed that our scheme can provide high security to the MECCO system while achieving significant performance improvement with minimum offloading and smart contract costs, compared to the existing schemes.   
\balance
\bibliography{Ref}
\bibliographystyle{IEEEtran}

\end{document}